# Design Representation as Semantic Networks

Serhad Sarica, Ji Han, Jianxi Luo

**Abstract**

Design representation is a common task in the design process to facilitate learning, analysis, redesign, communication, and other design activities. Traditional representation techniques rely on human expertise and manual construction and are difficult to repeat and scale. Here, we propose a methodology that utilizes a pre-trained large-scale cross-domain design knowledge base to automatically generate design representation as a semantic network, i.e., a network of the entities and relations, based on design descriptions in texts or natural languages. Our methodology requires no *ad hoc* statistics. Based on a participatory study, we reveal the effectiveness and differences of the semantic network representations that are automatically generated with alternative knowledge bases. The findings illuminate future research directions to enhance design representation as semantic networks.

## 1. Introduction

Design representation, i.e., representing design information or knowledge as notional objects, their relationships and structure in an abstract and graphical form, is a common and essential task in design processes (Shah and Wilson, 1989). Design representation facilitates designers' learning, analyses, communication, conceptualization, cognition and redesign of design artifacts (Cash and Maier, 2021). Conventional design representation techniques (examples in Fig. 1), such as sketch drawings (Yang and Cham, 2007) (Fig. 1A), design structure matrices (Eppinger and Browning, 2012) (Fig. 1B), functional diagrams (Stone and Wood, 2000) (Fig. 1C), and schematics (Rajan et al., 2005) (Fig. 1D), require domain knowledge of humans and manual construction. Design representation is considered a creative act, and the capability to do it well underlines human creativity (Oxman, 1997).

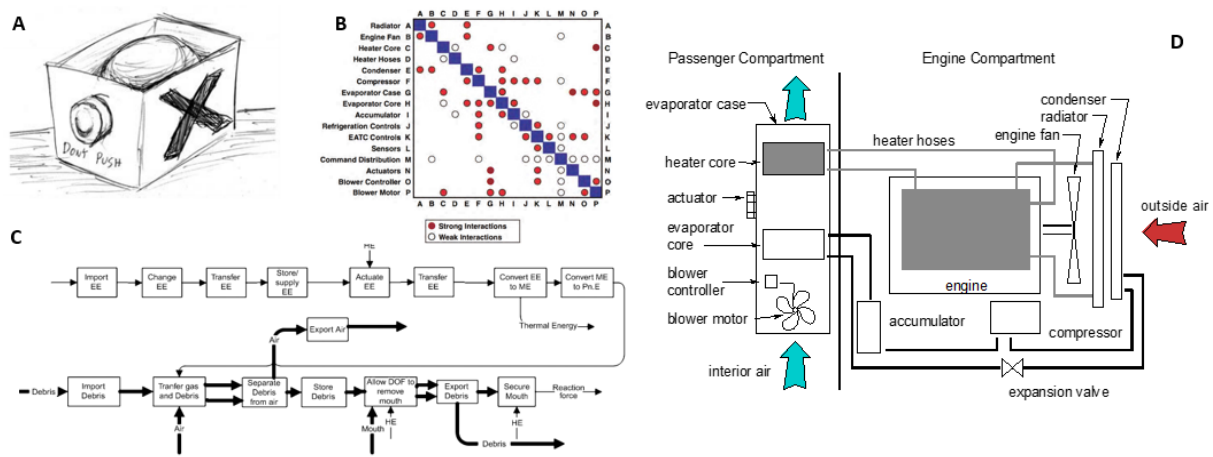

*Figure 1. Conventional design representation techniques: A) sketch drawings, B) design structure matrices, C) functional diagrams, D) schematics*

Automatic design representation is increasingly expected alongside the growing adoption of data science and artificial intelligence (AI) in design processes. In particular, the mass majority of design knowledge or information is actually stored in unstructured natural language documents or discourses (Ur-Rahman and Harding, 2012). Therefore, in this study, we focus on automating the generation of design representations based on design-related natural language descriptions. With natural language processing (NLP), our design representation refers specifically to design knowledge representation, rather than design embodiments such as sketches, CAD models, physical mockups, or prototypes.

In design research, various statistical techniques have been utilized to extract, map and analyze the topics within a collection of design documents and discourses (Chiarello et al., 2019; Dong, 2005; Dong et al., 2004; Dong and Agogino, 1996; Jiao and Qu, 2019; H. Song et al., 2020). However, analyzing a local document dataset and employing local association rules might not lead to a proper representation of the true and universal design-related associations of the design entities occurring in a document or dataset. A specific entity or entity-to-entity relation that is essential for a design topic may not occur frequently and thus may not be statistically significant in a short text, a single document or a set of a few documents that describe the design topic. A designer can read a document to recognize different design entities in the text and determine how they are associated if she or he possesses sufficient corresponding domain knowledge (Oxman, 1997). Nevertheless, such a human process to read design descriptions and create design representations is tedious, time-consuming, and prone to human errors or cognitive limitations.

The recent advancements in data science for engineering design (Chiarello et al., 2021) has allowed us to address these limitations. We propose a methodology that utilizes a pretrained large-scale cross-domain semantic network as the knowledge base to automatically generate a structural representation of the design knowledge by retrieving the design entities from a piece of design text or a document and their relations from the knowledge base. In so doing, the pretrained knowledge base serves as a *virtual multidisciplinary knowledge expert*. Several large public cross-domain knowledge bases, such as the common-sense WordNet (Miller et al., 1990) and ConceptNet (Speer and Lowry-Duda, 2017) and the engineering design focused B-Link (Shi et al., 2017) and TechNet (Sarica et al., 2020), are readily available to serve this purpose. These knowledge bases share one capability in common: they are very large semantic networks and enable the representation of a design description as a semantic network.

Although semantic network-based design representation is not new to the design literature (Han et al., 2022), our focus here is on its automatic generation by using a pre-trained knowledge base as a virtual expert instead of using human expertise or *ad hoc* statistics. In this regard, uncertainty and questions remain on the curation and choices of the pretrained knowledge bases for design representation in specific design contexts. To better understand the impact of knowledge bases on automatic design representations, we conducted a participatory experiment to investigate how well human designers can quickly comprehend the design information of the function, behavior, and structure of a multidisciplinary engineering design from its semantic network representation based on alternative knowledge bases.

The paper is organized as follows. We first review the relevant design research literature in Section 2. Section 3 describes the participatory study, and followed by discussions of the findings, implications and limitations in Section 4. Finally, the paper is concluded in Section 5.

## 2. Literature Review

Topic modelling methods along with related visualization extract the most prominent topics in a set of documents and latent semantic relations between the documents by utilizing methods such as latent semantic analysis (LSA) (Deerwester et al., 1990) and Latent Dirichlet Allocation (LDA) (Blei et al., 2003). Such methods have been adopted to support document and prior art search (Krestel and Smyth, 2013), generate relational design repositories (Fu et al., 2013), provide longitudinal changes in design domains (Chiarello et al., 2019), and assist innovative product design processes (Dong et al., 2004; Song et al., 2020). These methods provide rather coarse information about a set of documents and their contents. Additionally, when the text is short, there are few documents, and statistical significance is impossible, such local analytics are unable to capture the true or universal associations of the design entities described in the text documents.

Summarization and visual representation of design-related information or knowledge within a stand-alone design document or description, without *ad hoc* local statistics, would require readily available knowledge. Although a person can use his or her knowledge to rigorously read a design document or text description to discover all the design-related entities and determine their relations, such a cognitive process is tedious, labor-intensive, and requires the person to have relevant knowledge. Regarding descriptions of multidisciplinary designs or designs from domains other than the person's own expertise, he or she might not be able to create meaningful representations. Instead of relying on human expertise or *ad hoc* statistics, our strategy is to utilize a pretrained large semantic knowledge base to recognize the terms from a given text or discourse and retrieve their relations from the knowledge base automatically to represent the knowledge or text.

Several methodologies that build on predefined ontologies have been introduced to construct knowledge bases for specific design-related tasks and for particular design domains (Ahmed et al., 2007; Gero and Kannengiesser, 2014; Hu et al., 2017; Li and Ramani, 2007; Ming et al., 2020) , such as the Function-Behavior-Structure (FBS) (Gero, 1990; Goel et al., 2009) and State-Action-Part-Phenomenon-Input-oRgan-Effect (SAPPhIRE) (Chakrabarti et al., 2005). However, these studies require manual encoding based on domain knowledge and repetitive human efforts to populate the knowledge bases beyond the domain of interest. Hence, the adoption of these knowledge bases for unrelated domains is not feasible and realistic.

In the past two decades, several large public lexical databases, including WordNet (Miller et al., 1990), ConceptNet (Speer and Lowry-Duda, 2017), YAGO (Suchanek et al., 2007) and NELL (Mitchell et al., 2018), have gained growing popularity in various applications for knowledge discovery, question-answering systems, recommendation systems, and artificial intelligence (AI). These databases are normally comprised of a comprehensive collection of entities (e.g., words or phrases) and semantic relations among these entities, and are known as  knowledge graphs or semantic networks (Han et al., 2022). The entities and associations have been statistically or linguistically "trained" based on collaboratively edited and accumulated common-sense knowledge bases, such as DBPedia, Wikipedia and Wiktionary.

Thus, these large, cross-domain, pretrained knowledge bases may serve as the knowledge source for retrieving the essential entities and their relations from natural language design descriptions. Among them, WordNet and ConceptNet have already been employed in many design studies. WordNet is the earliest lexical digital database and is manually curated by experts through organizing words and phrases with same senses as synsets, which are interlinked by semantic and lexical relations such as synonymy, hyponymy and meronymy (Fellbaum, 2012; Miller et al., 1990). ConceptNet is a large knowledge graph (Speer and Lowry-Duda, 2017) automatically constructed using unsupervised learning algorithms to retrieve entities from Wikipedia, Wiktionary, WordNet, DBPedia and other resources and expert-created resources, connecting them with 34 types of relations, such as "IsA", "RelatedTo" and "PartOf".

Several studies have leveraged WordNet to analyze the semantic content in design processes. For example, Segers et al. (2005) used WordNet to capture the semantic associations between annotations to create "word graphs" to enhance the fluency of architectural design sessions. To support brainstorming sessions, Linsey et al. (2012) Linsey et al. (2012) leveraged the hierarchical structure of WordNet to populate a "word tree", in which the functional features of a given design problem are represented as verbs, to guide the exploration of analogical solutions. Taura et al. (2012) applied WordNet to associate the concepts generated in a computer simulation. Han et al., (2018) utilized WordNet and domain ontology for retrieving ideal CAD assembly models with specific functions to support the reuse of assembly models at the design stage. Kan and Gero (2018)Kan and Gero (2018) employed WordNet to construct a linkograph that characterizes the sequential segments in the process of a design session and measured the entropy of the linkograph. Besides, WordNet has often been used as a knowledge source to evaluate design ideas. For instance, Georgiev and Georgiev (2018) Georgiev and Georgiev (2018) introduced metrics based on WordNet to measure polysemy, creativity and divergence of new ideas. Goucher-Lambert and Cagan (2019) used WordNet to calculate the semantic similarity of crowdsourced ideas to categorize and use them as stimuli in design idea generation. Nomaguchi et al. (2019) assessed the novelty of functional combinations in concepts by using the semantic similarity measures derived from WordNet.

While WordNet provides lexical relations, ConceptNet relates entities through several specific semantic relations. In design research, (Yuan and Hsieh, 2015)Yuan and Hsieh (2015) leveraged the associations in ConceptNet to frame the creation process and discover service design insights. Han et al. (2018a) J. Han et al. (2018a) used ConceptNet to associate the design knowledge extracted from design websites to create a design knowledge base for combinational design. J. Han et al. (2018b)Han et al. (2018b) directly used ConceptNet as the knowledge base for query expansion and ontology construction to support analogical design. Chen and Krishnamurthy (2020) presented an interactive procedure that retrieves the terms in ConceptNet to aid and simulate designers in the mind-mapping process. Han et al. (2020) used ConceptNet as the knowledge base to evaluate the novelty of design ideas by measuring the semantic distance between the elemental concepts within the ideas. Chen et al. (2022) employed ConceptNet to generate cues in question forms to promote designers in breadth exploration during mind-mapping.

Pre-trained on common sense data sources and covering layman knowledge, WordNet and ConceptNet might be insufficient to capture contextualized design knowledge from the design language descriptions in architecture, engineering, etc. In contrast, B-link (Shi et al., 2017) and TechNet (Sarica et al., 2020) are two large knowledge bases in the form of semantic networks and are constructed based on engineering design data sources. B-link is constructed by retrieving technical terms from engineering-related academic papers and online design blogs and deriving relations among the terms occurring within a context window. B-Link was utilized by Chen et al. (2019) as the knowledge base for cross-domain knowledge associations to support design concept generation. On the other hand, TechNet utilized unstructured textual information in patents as the data source and employed extensive preprocessing and noise reduction followed by language model training to determine vectorized representations of over 4 million terms. TechNet has been used as the knowledge base to augment query expansion (Sarica et al., 2019), idea generation (Sarica et al., 2021) and evaluation (Han et al., 2020).

Table 1 summarizes the four pretrained knowledge bases that are most commonly used in design research. Despite the differences, they are all very large, cover diverse knowledge across many domains, and were pretrained for a wide range of NLP applications within or across domains, in contrast with the manually curated, small, domain-specific knowledge bases for specific design tasks or domains. Based on their data sources, WordNet and ConceptNet may best represent laypersons with extensive common-sense knowledge while B-Link and TechNet may better represent "super" engineering designers with broad knowledge in all technological domains.

*Table 1. Characteristics of the four knowledge bases most used in design research*

| Knowledge Bases | Knowledge Data Sources | Number of Entities | Number of Relationships |
|---|---|---|---|
| WordNet | Human lexicographers | 155,236 | $6.5 \times 10^5$ |
| ConceptNet | DBPedia, Wiktionary, WordNet, Open Mind Common Sense crowdsourcing, and Cyc | 516,782 | $1.3 \times 10^{11}$ |
| B-Link | ScienceDirect technical paper publication database (20 years) and design blogs | 536,507 | $3.7 \times 10^6$ |
| TechNet | Complete US patent database (40 years) | 4,038,924 | $8.5 \times 10^{12}$ |

Most importantly and related to our interest, the structures of these pretrained knowledge bases enable semantic network representations. However, it is still unclear how the differences of these knowledge bases might affect the semantic network representations resulting from them and human designers' perception of the design represented. To develop a nuanced understanding in this regard, we conducted a participatory experiment to investigate the differences and effectiveness of the semantic network representations of the same design, which were automatically generated based on alternative knowledge bases.

## 3. The Participatory Study

### 3.1 Semantic network representation of the design based on natural language text data

In this study, we use a text that describes the design of a "spherical robot" (Fig. 2) to generate semantic networks. Spherical robots are spherically-shaped robots that roll to move on a surface. It combines various mechanical, electromechanical and electronic components to achieve rolling motion and enable capabilities such as remote control, autonomous motion, and data collection with built-in sensors. It is noteworthy that the spherical robot design gained general popularity with BB-8's appearance in the Star Wars movies. Since the movies, several spherical robot toys have been sold in toy and electronics stores worldwide. We chose this design because, first, it combines knowledge elements from several technological domains, such as mechanical and electronics; and, second, designers without prior engineering domain knowledge can also potentially learn and understand its function, structure, and behavior after spending some time reading the detailed natural language description.

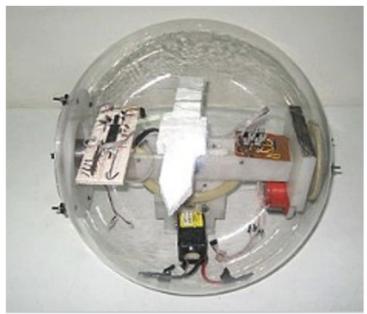

A Spherical Robot, also known as spherical mobile robot, or ball-shaped robot is a mobile robot with spherical external shape. A spherical robot is typically made of a spherical shell serving as the body of the robot and an internal driving unit (IDU) that enables the robot to move. Spherical mobile robots typically move by rolling over surfaces. The rolling motion is commonly performed by changing the robot's center of mass (i.e., pendulum-driven system), but there exist some other driving mechanisms. In a wider sense, however, the term "spherical robot" may also be referred to a stationary robot with two rotary joints and one prismatic joint which forms a spherical coordinate system (e.g., Stanford arm).

A pendulum-driven spherical mobile robot. (The white arrow is used to determine the position and orientation of the robot via a vision-based algorithm.

The spherical shell is usually made of solid transparent material but it can also be made of opaque or flexible material for special applications or because of special drive mechanisms.

The spherical shell can fully seal the robot from the outside environment. There exist reconfigurable spherical robots that can transform the spherical shell into other structures and perform other tasks aside from rolling.

Spherical robots can operate as autonomous robots, or as remotely controlled (teleoperated) robots. In almost all the spherical robots, communication between the internal driving unit and the external control unit (data logging or navigation system) is wireless because of the mobility and closed nature of the spherical shell. The power source of these robots is mostly a battery located inside the robot but there exist some spherical robots that utilize solar cells. Spherical mobile robots can be categorized either by their application or by their drive mechanism.

*Figure 2. Design description of "Spherical Robot" used in the case study (Source:*

*Wikipedia, accessed on June 8, 2020)*

With the design description in Fig. 2 as the input, we followed the procedure depicted in Fig. 3 to create a semantic network representation based on a knowledge base. The first step is to retrieve the terms (i.e. words and phrases) within the technical description which are already defined in the selected knowledge base. Second, the retrieved terms are associated with each other using the available pairwise associations of these terms within the knowledge base. In this study, we test three different knowledge databases, including WordNet, ConceptNet and TechNet, because they are publicly accessible via their APIs.

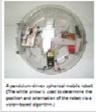 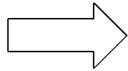 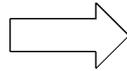 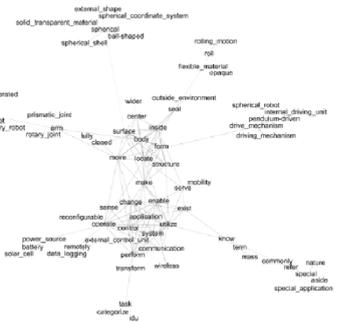

*Figure 3. The procedure to generate a semantic network from a design description*

We collected 75, 77, and 73 unique terms from the same text (in Fig. 2) based on WordNet, ConceptNet, and TechNet, respectively. Table 2 presents the numbers of retrieved unigrams, bigrams, and trigrams. We find that, different from WordNet and ConceptNet, TechNet allows the retrieval of longer and more detailed design-related entities such as "internal driving unit", "spherical coordinate system", and "solid transparent material".

*Table 2. Number of unigrams, bigrams and trigrams retrieved from the text using the lexicons of WordNet, ConceptNet, and TechNet, respectively*

|  | Unigram | Bigram | Trigram | Total |
|---|---|---|---|---|
| WordNet | 72 | 3 | - | 75 |
| ConceptNet | 72 | 5 | - | 77 |
| TechNet | 52 | 17 | 4 | 73 |

Then, we generated an adjacency matrix $A$ with a size of $N$, where $N$ is the size of the retrieved term set, and $A_{ij}$ corresponds to the semantic association of terms $i$ and $j$ in the corresponding knowledge base. All three knowledge bases have the capabilities to represent the quantitative semantic similarity between the terms contained within the knowledge base, despite their structural differences. When WordNet is employed, the shortest path that connects the terms in WordNet (Pedersen et al., 2004) is used as the quantitative measure of the strength of the link between the terms in the semantic network. For TechNet and ConceptNet, we used the cosine similarity of the vector representations of the terms in these respective pretrained knowledge bases. Therefore, the semantic networks based on the original data have weighted links.

To derive a visualization that can effectively represent the relational design knowledge within the text and to enable rapid comprehension within minutes, we employed a filtering method proposed by Hidalgo et al. (2007) and Yan and Luo (2017) to reduce the complexity of the network visualization. The network links are first filtered to reduce the graph to a maximum spanning tree (MST). An MST contains the minimum set of the strongest links in the network to keep the network fully connected. Thus, to connect $N$ nodes, $N$-$1$ links are required in the MST. Building upon the MST backbone, next strongest links are added to the network until we have 2$N$ links in the network. As the last step,  a force-directed algorithm called Force Atlas-2 (Jacomy et al., 2014) is employed to finalize the network visualization. Fig. 4 presents the resultant semantic networks. They are expected to represent the design knowledge structure of the spherical robot, as described in the natural language text in Fig. 2.

Before we invited participants to evaluate the generated semantic networks, we first investigated the details of these representations and discovered several differences. TechNet (Fig. 4A) clumped common terms such as "body", "form", "locate", and "move" in the center while organizing closely related terms together as peripheral cliques. These clusters surrounding the center can be reorganized and recombined to conceptualize the spherical robot design described in the text. For example, the design entities related to the shape of the spherical robot (i.e., spherical, ball-shaped, and spherical shell) are positioned very closely, as are those related to electrical power.

A

B

C

Figure 4. Semantic network representation of spherical robot design based on A) TechNet, B) WordNet, and C) ConceptNet

Conversely, WordNet (Fig. 4B) groups some key design aspects of spherical robots, such as "spherical", "autonomous", and "ball shaped", in the center of the network. Unlike TechNet, WordNet do not contain engineering and technology-related phrases that carry essential information regarding the design aspects. In addition, it cannot provide cohesive peripheral clusters as TechNet does. Lastly, ConceptNet (Fig. 4C) derives meaningful associations in the center of the network. However, it lacks the ability to reveal distinct conceptually related cliques.

*3.2 Experiment workflow*

We developed a web-based interface for participants to evaluate the effectiveness of the three semantic networks in summarizing and representing the design of the spherical robot. The web-based interface facilitates the following workflow.

1) On the first web page, participants are first asked to read the "spherical robot" design description (Fig. 2). Second, they are asked to summarize the design using at least 50 words based on their reading and understanding of the design. This summarization task is included to force the participants to go through the natural language design description thoroughly and ensure that they understand the details of the spherical robot design as much as they can before they see the semantic networks. Only after they submit the design summary are they directed to the second web page.

2) Then, three semantic networks in Fig. 4 are presented to the participants, which are randomly sorted for each participant. They are asked to evaluate the effectiveness and performance of the three networks in representing the specific design of "spherical robot". The question *how well does this graph represent the specific design of 'spherical robot'?* was given; and the respondents answered the question using a 5-point Likert scale, including "not representative", "slightly representative", "moderately representative", "very representative", and "strongly representative".

3) Finally, the participants are asked to choose the network visualization they consider the best and why. The question "*which of the three graphs is the best representation of the specific design of 'spherical robot'?*" was asked. The main motivation behind this question was to resolve possible situations in which participants score different visualizations equally. There is also an area for the participants to provide their comments.

4) After a participant submits his or her evaluations by clicking the submission button, the web interface is closed. The evaluations are sent to the research team.

We did not define any time limit for the tasks. However, before starting, the participants were informed about the typical duration of the study, which is about 10–15 minutes.

*3.3 Participants*

The selected participants were either professional engineers working in the industry or PhD students from Singapore University of Technology and Design and Nanyang Technological University who had formal engineering education. Overall, 56 participants completed the online participatory study. The primary author of this paper was able to perform informal post-experiment interviews with 25 of them. Three out of 25 participants revealed that they either answered at least one question wrongly due to comprehending it differently than expected, or they had just chosen the first graph representation that appeared since they were in a hurry. Therefore, the data of these three participants were omitted.

*3.4. Results*

As reported in Fig. 5, participants generally evaluated the TechNet-based representation with higher performance scores than the other two representations. The distribution for the TechNet-based representation is right-skewed and indicates that most participants find it effectively representative. The distributions for WordNet-based and ConceptNet-based representations are skewed to the left and suggest their inadequacy in representing the design. Fig. 6 reports the single choices of the best representations by the participants. The figure shows that most participants chose the TechNet-based semantic network as the best design representation of the spherical robot described in the given natural language text.

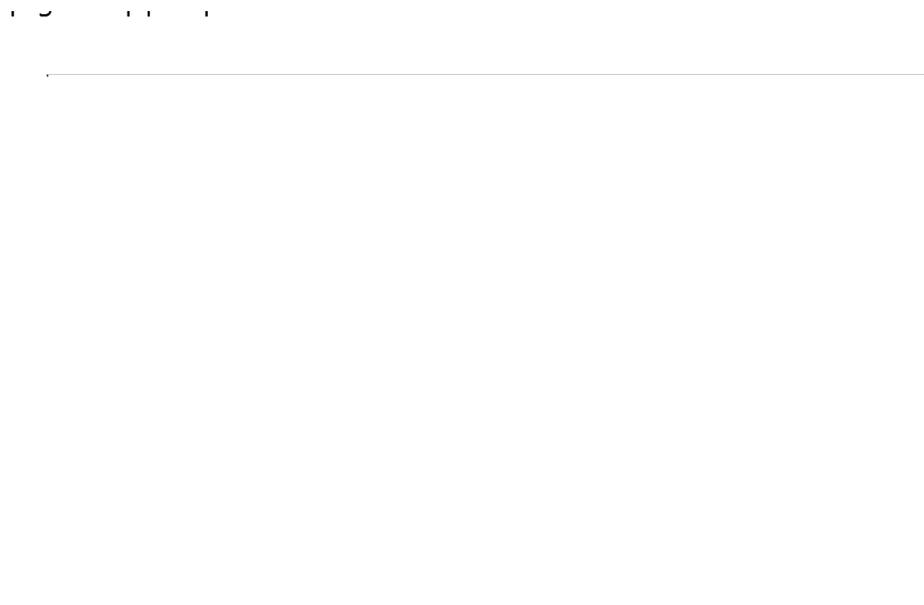

*Figure 5. Participants' evaluation of the design representation effectiveness*

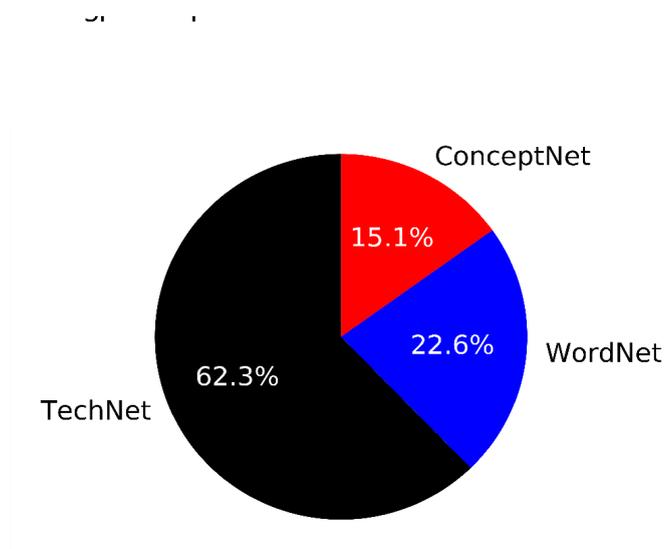

*Figure 6. Participants' single choice of the best representation.*

The main advantage of the TechNet-based network representation is that it can organize distinct groups of terms that are cohesive within and together constitute subsystems or abstract concepts that define the design under investigation. A few participants who chose TechNet representation as the best pointed out these specific characteristics. A participant commented the following: "*The nodes of the graph include multiword units (MWUs - e.g., spherical coordinate system) that are important for the reproduction of entities in the text description. The entities that comprise MWUs carry a specific meaning in this context, which is lost when these are decomposed into single word units*". Another participant also made a similar point by suggesting that *"B* (the TechNet-based representation in that case) *includes the specific concepts and components mentioned in the text while others draw a more general context. It is easier from B to understand that the graph is about spherical robots, while others do not possess the necessary information.*"

TechNet's ability to distinctly represent group subsystems as periphery clusters is also favored by some participants. One of them pointed out "*Visually it splits my attention into a few key components which we associate as different parts that can be used to describe a spherical robot*" while another participant commented *"It categorizes the keywords in some way and the categorization makes more sense than the others to me. Following the network graph, it's easier for me to think about which aspects I should consider for designing a spherical robot."*

The WordNet-based semantic network also presents a clear grouping of terms. However, it lacks design-specific terms that provide design-related clues. A participant who favored WordNet indicated its advantage by commenting that "*because it is the most organized and simple to use and study*" while another participant indicated that the "*arms are categorized in a more planned manner*".

On the other hand, sourcing semantic relations from the ConceptNet knowledge base results in a homogeneously drawn layout that lacks visibly distinctive groups. Nevertheless, some participants found the structure and the way ConceptNet connects the terms useful. For example, a participant ranked the ConceptNet representation as the best and indicated the following: "*It has a better connection between words as words which are related to each* *other are around each other. The other two graphs have multiple clusters which are connected to the center by a single connection which should not be right as all those words have relations with the words in the center*". Another participant focused on the center of the network and left the following comment: *"The links between the central design elements represent a more logical structure at graph C."*

The feedback and comments from the participants show that each of the network representations has some superior qualities over others whereas the survey statistics presented in Figs. 5 and 6 clearly suggest TechNet's advantage over others in the representation of the spherical robot design in this case, which is a technical engineering design.

## *4. Discussions*

### *4.1. Implications of the findings*

To this point, we have demonstrated the workflow of using a pretrained knowledge base as the virtual knowledge expert to automatically generate a semantic network-based design representation from a piece of natural language design text. From the participatory study, we also gained nuanced insights into the effectiveness and differences of alternative knowledge bases in generating design representations and inspirations. Such insights shed light on directions for future research and development.

### *1) Automatic design representation as a semantic network*

First, our design representation method distinguishes itself from the traditional ones in that it is automatic, flexible, and scalable. Classical design representation methods, such as the function diagram and design structure matrix, rely on human expertise to create the representation. Analyzing a piece of local text or a local set of documents cannot provide the statistical significance required to derive ground-truth relations between design elements. Existing domain-specific knowledge bases require human expertise to encode individual designs according to predefined ontologies for use in specific tasks and domains, and are inflexible and unscalable. Prior design studies leveraging large cross-domain pretrained knowledge bases normally adopted a single knowledge base for a specific design task.

For instance, WordTree (Linsey et al., 2012) is limited to WordNet, functional verbs, the tree topology, and the specific design task of idea stimulation, whereas our representation method is flexible regarding the choice of the knowledge bases, the terms in the graph, the graph's topologies, and the application tasks in the design process. Fig. 4 shows the automatically generated different representations with three alternative knowledge bases with the same methodological procedure (described in Section 3.1). These representations contain diverse terms that represent not only functions but also components (e.g., "solar cell" and "internal driving unit"), structures (e.g., "ball-shaped"), materials (e.g., "solid transparent material"), properties (e.g., "autonomous"), and working mechanisms (e.g., "pendulum-driven"). Our method allows flexible graph topologies to emerge in an unsupervised manner, such as the multiple modules with a core in Fig. 4A (TechNet) and 4B (WordNet) or the rather homogenous network structure in Fig. 4C (ConceptNet). In turn, such holistic, informative, and flexible representations can have various applications, such as communicating the design, facilitating the learning and analysis of the design, inspiring redesign, and other applications, in the design process.

Although our participatory study focused on representing a short text that describes the specific detailed design artifact of "spherical robots", our method is scalable and can be used to automatically process the natural language texts that describe more complex designs, such as an aircraft design or a shopping mall design; and that are either part of a document, an entire document such as a report, paper, patent, user manual, design review, or a collection of documents. Our method can also be used to extract and represent the design knowledge or information in natural language discourses, conversations, or communications in design processes or in the documents about design processes or activities in addition to artifacts.

Furthermore, the process to extract the terms and their relations from the pretrained semantic network knowledge bases does not require statistics and thus is computationally light. WordNet, ConceptNet, TechNet and other knowledge bases have already been pre-trained and ready for retrieval uses, although the processes to pretrain them were computationally expensive and sophisticated.

*2) Common-sense versus domain-specific knowledge bases*

Our second contribution lies in the comparative investigation of the effectiveness and differences of design representations drawn from alternative pretrained knowledge bases. While pretrained knowledge bases have already been used in different design tasks, individual studies normally make an arbitrary choice of knowledge bases without comparison and justification. Our participatory study shows that TechNet provides a more comprehensible, detailed, informative, and structured representation of the technical design knowledge of spherical robots. While WordNet and ConceptNet also provide informative representations of spherical robot designs, they are more abstract, general, and common-sense. One can view TechNet as a virtual expert with extensive multidisciplinary engineering design knowledge while WordNet and ConceptNet have rich common-sense knowledge.

The key learning point here is that while common-sense knowledge bases, such as WordNet and ConceptNet, have been most popularly employed in design studies, design representations may be enhanced by using more contextualized knowledge bases. For instance, engineering design (e.g., spherical robots) representations may benefit from an engineering knowledge base (e.g., TechNet), architecture design representation may benefit from an architectural knowledge base, and medical design representation may benefit from a medical knowledge base. At the same time, such a contextualized design knowledge base also should not be too narrow or specialized and thus be limited in its scalability for applications across domains and extensibility for different tasks. For example, a specialized mechanical design knowledge base will be inadequate to represent the spherical robot design and other multidisciplinary designs such as an autonomous electric vehicle.

It can be challenging for a specialized or inexperienced designer to comprehend information-heavy domain-specific terms from distant domains in a broad cross-domain semantic network. In such cases, the aid from a broader common-sense knowledge base will be helpful. Comprehending terms and associations from the common-sense WordNet and ConceptNet could be relatively easier. However, they might not provide sufficiently detailed and nuanced knowledge for a specialized domain to be useful for domain experts.

Therefore, the ideal knowledge base for design representation should integrate both common-sense knowledge and professional specialized design knowledge in engineering, architecture, medical, and/or more professional domains (as depicted in Figure 7). Such an integrated knowledge base may be capable of generating representations that make sense for designers with different knowledge breadths and depths, including laypersons, novice designers and domain experts. However, such an ideal knowledge base does not yet exist.

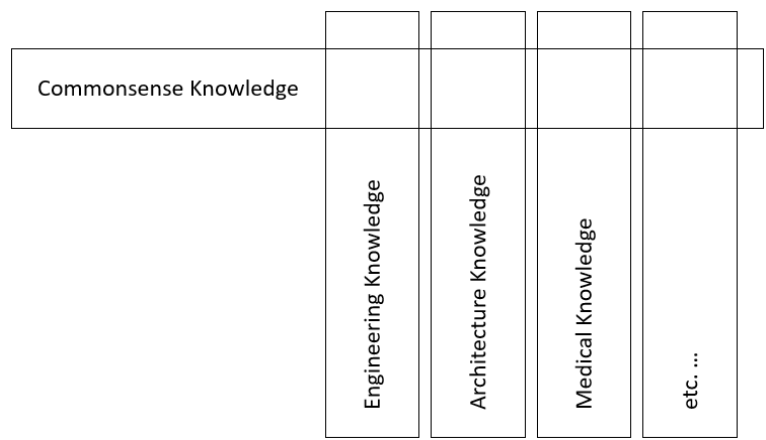

*Figure 7. Architecture of an ideal knowledge base for contextualized specialized design representation and flexible multidisciplinary design representation*

To date, most design studies still employ the two common-sense knowledge bases, WordNet and ConceptNet. In addition to the recently created and publicly available engineering knowledge base TechNet and B-Link, we also call for the construction of large knowledge bases (e.g., on the scale of TechNet) for architecture, medical and other generally defined fields of professional practices based on the knowledge data sources from those domains and blending their uses with the readily available common-sense knowledge bases. In the long run, we anticipate the development of an integrated knowledge base that encapsulates both extensive common-sense knowledge and deep domain knowledge from a variety of professional domains.

*3) Knowledge graph representation*

In the semantic network representations in Fig. 4, the entities are connected to one another with weighted links representing their quantitative semantic similarity or distance in respective knowledge bases. Such weighted linkage information enables network representations that are more informative than traditional topic maps and word clouds. The network representations can be further enhanced if the relations can explicitly inform how entities could be connected in the actual design and be more sensible for humans. In the field of artificial intelligence, qualitative networks composed of many heterogenous <entity, relation, entity> triplets are specifically referred to as knowledge graphs (Amit Singhal, 2012; Ehrlinger and Wöß, 2016).

A knowledge graph is a special type of semantic network. Among the large cross-domain knowledge databases used in prior design studies, only ConceptNet qualifies as a knowledge graph. It consists of 34 types of predefined relations between entities, such as "IsA", "RelatedTo", "CapableOf" and "AtLocation", which can be understood easily by both computers and humans. Fig. 8A shows the ConceptNet-based knowledge graph presentation of a sentence from the text describing spherical robots in Fig. 2. The relations covered are of common sense and provide little value to inform design. For a knowledge graph to effectively represent designs, one would expect more diverse <entity, relationship, entity> triplets that carry more specific design information, such as <power source, be, battery> and <spherical robot, utilize, solar cell>, as exemplified in Fig. 8B.

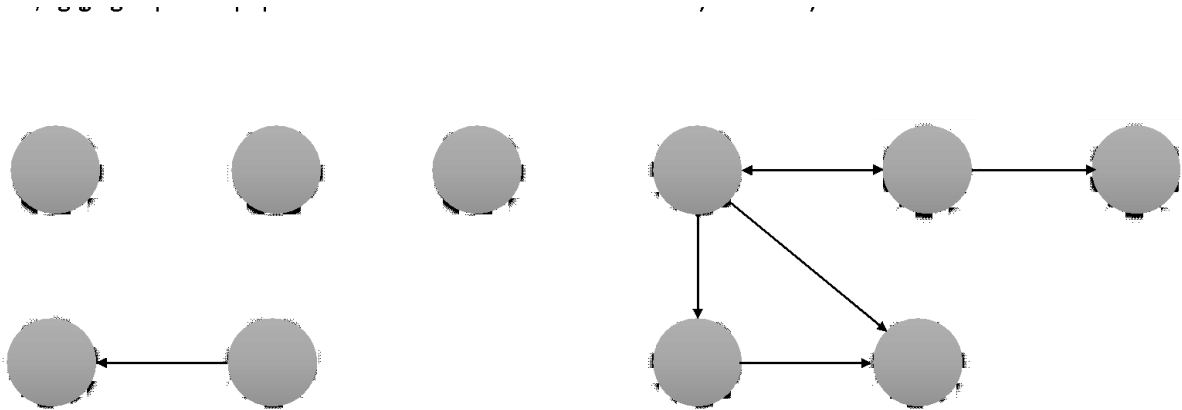

*Figure 8. Design knowledge graph versus common-sense knowledge graph.*

Therefore, a "design knowledge graph" may enable more informative and comprehensible design knowledge representation than a commonsense knowledge graph such as ConceptNet or a quantitative semantic network such as TechNet. To date, a "design knowledge graph" at the scale and scope of ConceptNet or TechNet is not publicly available. In addition to our own ongoing work to develop a large-scale multidisciplinary engineering design knowledge graph, recently a few other scholars have also explored different techniques to represent design knowledge as knowledge graphs based on patent data (Deng et al., 2019; Siddharth et al., 2022; Zuo et al., 2021). However, none of these efforts has implemented public APIs or interfaces for external researchers and communities to access their knowledge graphs.

Taken together, our final proposition from this study is to develop a large-scale cross-domain semantic network knowledge base in which both quantitative and qualitative relations between entities are pretrained with both common-sense knowledge and engineering, architecture, medical and other domain knowledge. While quantitative relations (e.g., semantic similarity or distance) are useful to determine the structure and topology of the semantic network representation, qualitative relational information is more sensible for human comprehension. While domain knowledge provides more nuanced and detailed design information useful for domain experts and specialized designers, common-sense knowledge helps relate the design to users and stakeholders that are laypersons or experts in other domains. Quantitative and qualitative relational information, common-sense and domain knowledge can be blended to generate design representations for a variety of goals and users in context.

*4.2. Limitations*

There are a few limitations with this study that need to be addressed in future work. First of all, although the introduced methodology for producing network visualizations of design descriptions using semantic networks is generic and built on well-established layout and filtering methods, it is specific. There are various other filtering approaches and network layout algorithms that can be adopted and tested. However, covering various methodologies may introduce challenges in the evaluation of alternative knowledge bases. Second, there are many alternative approaches to network visualizations to create visual summaries of a piece of text. Engineering design scholars frequently used word clouds or hierarchical layouts such as mind maps to generate representations. Third, other potentially useful knowledge bases other than those included in the present study, especially those that have yet to be utilized in the design research literature, can be further explored. Fourth, using only textual data to represent a design could limit the users in comprehending, learning, and conceptualizing the design knowledge. Mixed uses of textual data and pictorial data (such as sketches, CAD models and images) for design representation could be further explored.

## 5. Concluding Remarks

In this study, we propose to automate design representation based on natural language design data as semantic networks specifically by employing pretrained, large-scale, cross-domain semantic knowledge bases as virtual knowledge experts rather than human expertise or local document statistics.    Our participatory study demonstrates the proposed methodology using three alternative knowledge bases and reveals the differences in the generated design representations.  Specifically, the TechNet-based semantic network provides more design-related details than the more widely used common-sense knowledge bases WordNet and ConceptNet. The findings suggest that the integrated or blended uses of common-sense and design domain-specific knowledge bases may yield balanced knowledge representations for the comprehension of laypersons, novice designers and domain experts.

Overall, this study is only a first step in automating design representations with pretrained knowledge bases. It provides a novel and effective approach for design representations as semantic networks, which has presented a contribution to the knowledge in the intersection of engineering design and data science. The study should be viewed as an invitation for further research, methodological improvements, and applications that leverage large pre-built knowledge bases to represent complex and multidisciplinary designs automatically, efficiently, and accurately from unstructured natural language texts, documents, and discourses.